# High-Field Terahertz Spin Resonance in $Cr_2O_3$ above the Spin-Flop Transition

Kaiyang Huang, Yuto Kinoshita[*], Natsuki Kanda[†], Takuya Matsuda[#], Masashi Tokunaga, Ryusuke Matsunaga, Yasuhiro H. Matsuda[*]

*Institute for Solid State Physics, The University of Tokyo, Kashiwa, Chiba 277-8581, Japan*

**Abstract** We report single-shot terahertz time-domain spectroscopy of $Cr_2O_3$ in pulsed magnetic fields up to 30 T. Well above the spin-flop field, in the 20-30 T range, the resonance frequency exhibits a nearly linear field dependence with a slope of ~22 GHz/T, smaller than the 28 GHz/T reported from low-field measurements. This reduction is insensitive to temperature and to a 15° field tilt, suggesting an intrinsic high-field property.

$Cr_2O_3$ has a corundum-type structure and exhibits antiferromagnetic (AFM) order with spins aligned along the *c* axis below $T_N$ = 307 K.[1)] This AFM order breaks both the space-inversion and time-reversal symmetries, resulting in a finite linear magnetoelectric effect.[2,3)] Below the spin-flop transition, two branches of antiferromagnetic resonance (AFMR) are present. At zero magnetic field, these modes are degenerate at a resonance frequency of 0.16 THz, and the degeneracy is lifted upon application of a magnetic field along the *c* axis.[4,5)] Above approximately 6 T, a spin-flop transition occurs, in which the spins align nearly perpendicular to the magnetic field. In this phase, a quasi-ferromagnetic resonance (QFMR) mode emerges, corresponding to the precession of the net sublattice magnetization about the external magnetic field.[5-7)] Although the high-field behavior of this mode has been discussed theoretically, experimental information in the THz range above 10 T has remained limited.[5,7)]

We studied this regime using a low-temperature terahertz time-domain spectroscopy (THz-TDS) system combined with a 30 T mini-coil pulsed magnet at the Institute for Solid State Physics, The University of Tokyo. The mini-coil generates pulsed magnetic fields via a current discharged from a compact capacitor bank (capacitance of 1.2 mF), producing magnetic fields up to 30 T with a pulse width of 1 ms at a charging voltage of 1900 V. The experimental setup is based on a single-shot THz-TDS technique for pulsed high magnetic fields,[8)] following the implementation established by the Rice group for the 30 T Rice Advanced Magnet with Broadband Optics (RAMBO) system.[9)] A laser pulse from a Ti:sapphire laser amplifier (central wavelength of 800 nm, repetition rate of 1 kHz, and pulse energy of ~1 mJ) was split into two beams: one serving as a pump pulse to generate a THz pulse via a $LiNbO_3$ crystal using the tilted-pulse-front method [10,11)] and the other serving as a probe pulse to detect the transmitted THz pulse using an electro-optic (EO) sampling technique with a (110)-oriented ZnTe crystal. The generated THz pulse was focused onto the sample using an off-axis parabolic mirror, where it was linearly polarized in the plane perpendicular to the magnetic field by a wire grid polarizer, and the transmitted THz pulse was subsequently focused onto the EO crystal. Single-shot detection scheme using a reflective echelon mirror and a CMOS camera was employed with a time window of approximately 20 ps. The pulsed-magnetic field system was synchronized with the laser amplifier to probe a spin resonance under high magnetic fields. We recorded the transmitted THz waveform with and without the magnetic-field pulse (Fig. 1(a)). By subtracting the zero-field reference from the finite-field signal, we extracted the field-induced change in the THz time-domain response of the sample, as shown in Fig. 1(b). The sample was mounted at the center of the coil using a sapphire holder extending from the cold stage. The temperature was controlled down to 13 K using a cryogen-free closed-cycle helium refrigerator, which cools both the mini-coil and the sample, with a base temperature of 10 K.

The sample is a commercially available $c$-plane single crystal grown by the Verneuil method, with a thickness of approximately 0.5 mm. The magnetic field is applied parallel to the $c$ axis. The blue and red curves in Fig. 1(a) represent the time-domain waveforms of the transmitted THz pulse at the peak of the applied pulsed magnetic field (28.8 T) and just before the field is applied, respectively. The temperature is 78 K. The single-cycle waveform at 0 ps corresponds to the THz pulse transmitted through the sample, while the smaller single-cycle component at 11 ps originates from a THz pulse reflected at the back surface of the sample. Under the magnetic field, a clear oscillatory structure is observed. By taking the Fourier transform of the differential waveform shown in Fig. 1(b), a resonance spectrum with a peak around 0.68 THz is obtained, as indicated by the pink curve in Fig. 1(c). By varying the magnitude of the applied pulsed magnetic field, the resonance spectra at each field are obtained, as shown in Fig. 1(c). Above 15.4 T, a clear resonance peak appears and exhibits a blueshift with increasing magnetic field. Measurements were performed at various temperatures, including 15, 78, and 283 K. In addition, separate measurements were carried out at 277 K with the sample tilted by approximately 15° relative to the magnetic field. The resonance peak frequencies ($\omega/2\pi$) are plotted as a function of magnetic field ($H$) in Fig. 1(d). It is found that the $\omega$-$H$ relation is nearly linear above 20 T. Above the spin-flop transition, the $\omega$-$H$ relation of the QFMR is given by

$$\frac{\omega}{2\pi} = \gamma\sqrt{H^2 - 2H_a H_e}\ , (1)$$

where $H$ is the external field, $H_a$ the anisotropy field, and $H_e$ the exchange field, and $\gamma$ is the gyromagnetic ratio.[5] Using the reported values of $H_a = 0.07$ T and $H_e = 245$ T (with $2H_a H_e \approx 36$ T$^2$) at low temperature,[5] and $\gamma = 28$ GHz/T, the calculated result is shown as the red line in Fig. 1(d). In the high-field limit $H^2 \gg 2H_a H_e$, this expression reduces to $\omega \approx \gamma H$, and the slope of the linear $\omega$-$H$ relation directly reflects $\gamma$. However, it deviates from the experimental results. Although the $H_a H_e$ term is temperature dependent near room temperature,[5] it becomes negligible in high magnetic fields, and thus the absence of temperature dependence in the high-field regime is expected. For a tilted sample, Foner et al. reported a shift of the resonance frequency when the field is tilted away from the easy axis near the spin-flop regime.[5] By contrast, in the present 20–30 T range well above the spin-flop transition, no clear change is observed within our resolution. To quantify the high-field slope, we fitted the data in the linear range of 20-30 T and obtained an effective slope of about 22 GHz/T, as shown in Fig. 1(e). The slope corresponds to the gyromagnetic ratio $\gamma$. The fitted value ($\gamma = 22$ GHz/T) is about 78% of the nominal 28 GHz/T (corresponding to $g = 2.00$, where $g$ is the $g$-value for Cr ions). In addition, the resonance intensity is markedly enhanced above 20 T, as shown in Fig. 1(c). This enhancement is attributed to the increase in magnetization along the magnetic-field direction with increasing applied field.

Several mechanisms may contribute to this discrepancy in $\gamma$. The simplest possibility—that the applied field is smaller than expected—can be ruled out, as direct measurements using a pickup coil at the sample position confirm that the expected field is applied. Among other possible causes, magnetoelastic renormalization of the anisotropy in the spin-flop phase appears to be the most plausible explanation. In antiferromagnets, magnetostriction is, to leading order, quadratic in the magnetic field. Through magnetoelastic coupling,[12,13] this can generate an effective anisotropy of the form $H_{a,\mathrm{eff}} = H_{a0} + bH^2$. Substituting this into Eq. (1) yields

$$\frac{\omega}{2\pi} = \gamma\sqrt{(1 - 2H_e b)H^2 - 2H_{a0}H_e}, \quad (2)$$

Accordingly, the high-field slope is reduced to $\gamma_{eff} = \gamma\sqrt{1 - 2bH_e}$ for $b > 0$. Similar field-dependent anisotropy or resonance renormalization attributed to magnetoelastic effects has been reported in several antiferromagnets, including $LiFePO_4$, $MnCl_2$, and α-$Fe_2O_3$.[13–15] To compare quantities with the same dimension, we estimate $\Delta H_a = bH^2$. With $b = 3.8 \times 10^{-4}$ T$^{-1}$, $\Delta H_a$ is approximately 0.15–0.34 T in the 20–30 T range, comparable in magnitude to the monodomainization fields of order 0.2–1.71 T reported for several antiferromagnets in Ref. 16. This order-of-magnitude agreement supports the plausibility of the required magnetoelastic anisotropy correction. In addition, $Cr_2O_3$ itself is known to exhibit abrupt length changes near the spin-flop transition and a strong pressure dependence of its anisotropy-related critical field, supporting the possibility that field-induced strain feeds back on the anisotropy.[17-19] From the experimental results, the value of $b$ is estimated to be $3.8\times10^{-4}$ T$^{-1}$. This value is obtained by fitting the data with Eq. (2), and the corresponding fitted curve is shown as the dashed line in Fig. 1(d). We note that this modified calculation is introduced as a phenomenological correction for the high-field spin-flop phase and is not intended to redetermine the spin-flop field. (The high field approximation is less accurate near the spin-flop transition field.) To verify this scenario, further investigations such as magnetostriction measurements under high magnetic fields on the same sample are required.

**Acknowledgements**

This work was supported by JSPS KAKENHI Grant Numbers JP20K15159, JP23K25814, JP23H04860, and the Daiichi-Sankyo "Habataku" Support Program for the Next Generation of Researchers.

*Corresponding author. E-mail:kinoshita@issp.u-tokyo.ac.jp, ymatsuda@issp.u-tokyo.ac.jp

†Present address: Extreme Photonics Research Team, RIKEN Center for Advanced Photonics, RIKEN, Wako, Saitama 351-0198, Japan

#Present address: Department of Physics, Graduate School of Science, Osaka Metropolitan University, Sumiyoshi-ku, Osaka 558-8585, Japan

Ref:

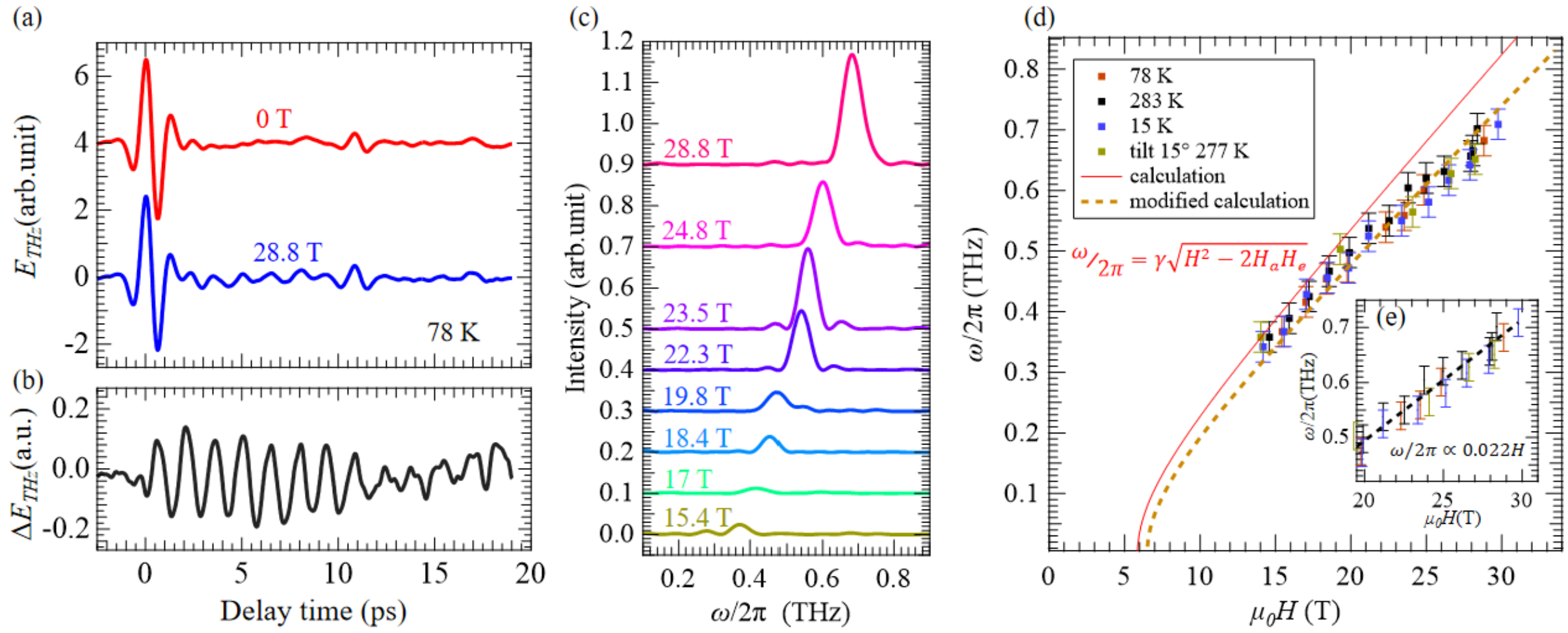

Fig. 1(Color online) (a) Time-domain waveforms of the transmitted THz pulse at 28.8 T and 0 T at 78 K. The magnetic field was applied along the *c* axis. (b) Differential waveform between the 28.8 T and 0 T signals. (c) Resonance spectra at various magnetic fields at 78 K. (d) *ω*–*H* relationship obtained under several measurement conditions. (e) Linear fit in the high-field regime.